\documentclass[letterpaper,journal]{IEEEtran}
\usepackage{amsmath,amsfonts}
\usepackage{array}
\usepackage[caption=false,font=normalsize,labelfont=sf,textfont=sf]{subfig}
\usepackage{textcomp}
\usepackage{stfloats}
\usepackage{hyperref}
\usepackage{verbatim}
\usepackage{graphicx}
\usepackage{cite}
\usepackage[T1]{fontenc}

\usepackage{amssymb,amsthm}

\usepackage{booktabs,multirow}

\begin{document}
\raggedbottom
\title{Physics-Informed Koopman Learning for Data-Efficient System Identification of Automated Vehicle Dynamics}
\author{Hangyu Li$^{1}$, Yiming Meng$^{2,*}$, Zhaohui Liang$^{1}$, Yunqing Xiao$^{3}$, Yuxuan Chen$^{3}$, Bofeng Cao$^{1}$, Shixiao Liang$^{1}$, Jiaxi Liu$^{1}$, Haotian Shi$^{1}$, Xiaopeng Li$^{1}$
\thanks{$^*$ Corresponding author: Yiming Meng (e-mail: yimingmeng@hkust-gz.edu.cn).}
\thanks{$^{1}$ Hangyu Li, Zhaohui Liang, Bofeng Cao, Shixiao Liang, Jiaxi Liu, Haotian Shi, and Xiaopeng Li were with the Department of Civil and Environmental Engineering, University of Wisconsin-Madison, Madison, Wisconsin, 53706, United States.}
\thanks{$^{2}$ Yiming Meng was with the Coordinated Science Laboratory, University of Illinois Urbana-Champaign, Urbana, Illinois, 61801, United States.}
\thanks{$^{3}$ Yuxuan Chen and Yunqing Xiao were with the Department of Electrical and Computer Engineering, University of Wisconsin-Madison, Madison, Wisconsin, 53706, United States.}}
\maketitle

\begin{abstract}
Understanding how a particular automated vehicle reacts to nearby traffic can improve motion prediction, control development, traffic simulation, and safety evaluation. However, commercial automated driving systems are proprietary, and single encounters may provide only a few seconds of noisy data. Standard physical models are interpretable but inflexible, whereas neural models are often data-hungry. We propose a physics-informed Koopman framework that combines a calibrated car-following law with a correction learned from the observations. A weak-form estimator learns this correction from velocity measurements integrated over time, avoiding noise-sensitive numerical differentiation. The resulting continuous-time model predicts the follower's response and provides interpretable measures of driving behavior. Experiments on OpenACC and CATS-CF evaluate 5-, 7-, and 11-second histories. Averaged over these histories, the constant-time-headway branch reduces speed prediction error by 16.0\% and 15.4\% on OpenACC and CATS-CF, respectively, relative to the best neural baselines. The PI-Koopman branches achieve the lowest mean speed and gap errors among the physical and neural comparators. Weak-form estimation reduces acceleration-field reconstruction error by 74.5\% in noisy known-law simulations and noise-induced field perturbation by 91.6\% in paired real-vehicle measurements. The same PI-Koopman system identification supports following-mode recognition with an approximately eleven-fold difference in median residual variance between human drivers and automated vehicles, and a single residual threshold achieves 85.4\% balanced accuracy. The branch informed by the intelligent driver model takes a median 17 to 26 ms for identification and forecasting on the tested central processing units, supporting frequent target adaptation. The CATS-CF dataset and code are available at \hyperref[]{https://github.com/CATS-Lab/PI-Koopman}.
\end{abstract}


\section{Introduction}
\IEEEPARstart{A}{utomated} vehicles respond to surrounding traffic through control strategies that are generally inaccessible to external observers. A recorded trajectory describes how a vehicle moved during an encounter, whereas a response law characterizes how it reacts to the state and motion of surrounding road users. Such models can inform motion prediction, interaction planning, traffic simulation, and safety evaluation~\cite{mahbub2023safety,li2025optimized}. Yet a moving observer or roadside unit may observe a particular vehicle for only a brief encounter~\cite{jiang2021deployment,li2024deception,ma2026real}. The system identification task is therefore to infer a queryable model of the vehicle's response from a short history.

Commercial adaptive cruise control (ACC) provides a production realization of automated longitudinal control, with proprietary strategies and responses that vary across vehicles and operating conditions~\cite{gunter2019model,gunter2020commercially,yu2025dynamic,li2025robotic}. Physical car-following laws provide interpretable response structure and can be calibrated to individual vehicles~\cite{wang2020online,wang2022identifiability}, but their prescribed functional forms may not capture all vehicle-specific response variations. Neural network models offer greater flexibility, yet a short encounter provides limited excitation for estimating a large number of parameters. Numerical differentiation of noisy velocity can amplify errors in acceleration estimation, leading to another difficulty~\cite{schaeffer2017sparse,bortz2023direct}. These challenges call for a system identification framework that combines physical structure, a compact adaptable representation, and noise-robust estimation.

We therefore propose a physics-informed Koopman learning (PI-Koopman) framework to identify a compact continuous-time response representation~\cite{mezic2021koopman}. The Koopman generator acting on the follower-state observable derives the acceleration under a specified car-following input~\cite{mauroy2019koopman,klus2020data,proctor2018generalizing}. A constant-time-headway (CTH)~\cite{swaroop1994comparision} or intelligent driver model (IDM)~\cite{treiber2000congested} law is first calibrated to the history states, after which a compact dictionary correction captures deviations from the nominal physical model. The correction provides flexibility without discarding the calibrated prior. Rather than relying on acceleration estimates obtained through numerical differentiation, weak temporal equations learn the correction from integrated velocity observations and dictionary terms.

Performance is evaluated through real-vehicle datasets. Among paired-vehicle datasets~\cite{xu2023v2v4real,li2026cats}, OpenACC provides production automated vehicle variation for evaluating system identification errors in a large-scale generalized environment~\cite{makridis2021openacc}. As a supplement, we collected 42 more repeated car-following data pieces using two same-model vehicles, with two ACC settings and two human drivers, forming our CATS-CF dataset. Synchronized and RTK-fixed 125 Hz 30-second GNSS/INS records demonstrate high-precision vehicle locations and motions, allowing evaluations of the proposed PI-Koopman framework under different sampling rates and history lengths. Given the high-precision nature of the data, we also synthesize different levels of errors into the data to demonstrate the stabilization of the PI-Koopman framework to noise. Together, these experiments thoroughly evaluate estimation accuracy, generalization, and noise-robustness of the system identification model.

Beyond motion prediction, the identified response provides a direct way to recognize ACC from traffic flow. Under a fixed fitting protocol, residuals quantify how consistently one response law explains an observed encounter. Our repeated CATS-CF trials link this quantity to ACC and human labels. The full PI-Koopman response is fitted independently to each encounter, and the variability of its velocity-integral residuals supplies a single ACC recognition feature. Human drivers have approximately eleven times the median residual variance of ACC encounters. A threshold learned from training sets achieves 85.4\% balanced accuracy on testing runs, demonstrating an additional use of the system identification method.

In general, this study makes three main contributions:
\begin{enumerate}
\item \emph{Data-efficient ACC system identification.} Across histories of 5, 7, and 11~s, PI-Koopman-CTH achieves the lowest mean speed errors on OpenACC and CATS-CF, reducing speed RMSE by 16.0\% and 15.4\%, respectively, relative to the best neural baselines. The CTH and IDM branches achieve the lowest mean gap errors on OpenACC and CATS-CF, respectively.
\item \emph{Stable recovery of automated vehicle dynamics.} Within a common dictionary, weak estimation reduces noisy acceleration-field reconstruction error by 74.5\% relative to finite differences at 25 Hz. Under synthetic errors in the 125-Hz real-vehicle data, it reduces response-field perturbation by 91.6\%.
\item \emph{Response consistency for ACC recognition.} PI-Koopman residual variability distinguishes ACC with 85.4\% balanced accuracy using a single threshold. The demonstration of one potential downstream application extends the model from prediction to behavioral characterization.
\end{enumerate}

\section{Related Work}
\subsection{Physical structure and short-encounter identification}
Calibrated car-following laws characterize vehicle responses and disturbance propagation~\cite{milanes2014modeling,gunter2019model}. Online parameter estimation and practical identifiability establish the role of excitation and observation history~\cite{wang2020online,wang2022identifiability}. Asymmetric ACC laws and extensions of IDM capture acceleration limits and additional response mechanisms~\cite{zhou2022empirical,shang2024two,bai2025novel}. These models provide useful response priors for encounter-specific adaptation.

Physics-informed learning and kinematics-aware prediction combine prescribed structure with learned response variation~\cite{mo2021physics,sheng2024kinematics}. Physics-enhanced residual learning (PERL) couples a calibrated physical predictor with a neural correction for data-efficient vehicle trajectory prediction~\cite{long2025physical,liang2025theory}. PI-Koopman extends this principle from trajectory-level correction to continuous-time response identification. A compact dictionary and weak velocity equations identify a correction to the acceleration field, conditioned on measured leader motion.

Neural trajectory models also use information accumulated across encounters. TransFollower learns long-sequence car-following prediction from a large collection of driving events~\cite{zhu2022transfollower}, while PreTraM improves trajectory-data efficiency through trajectory-map pretraining~\cite{xu2022pretram}. This distinction between training exposure and the observation history at inference motivates our matched target-history comparison. Source groups select model configurations, and a short encounter supplies the target-specific fit.

\subsection{Koopman representations and temporal estimation}
Extended dynamic mode decomposition represents observable evolution through a finite dictionary~\cite{williams2015data}. Generator-based identification relates observable evolution to nonlinear vector fields~\cite{mauroy2019koopman,klus2020data}, and explicit input formulations extend Koopman modeling to forced systems~\cite{proctor2018generalizing}. Learned embeddings use neural encoders to obtain coordinates with linear latent evolution~\cite{lusch2018deep}. Traffic applications include platoon prediction, follower-acceleration reconstruction, and mixed-traffic control~\cite{tian2025physically,mustavee2025koopman,lyu2026mitigating}. Adaptive deep Koopman models combine acceleration-informed losses with online operator updates~\cite{zhang2026physics}, while residual Koopman control learns compensation for a physical control input~\cite{fu2025residual}.

Integral regression, weak SINDy, and WENDy estimate differential equations through temporal integration~\cite{schaeffer2017sparse,messenger2021weak,bortz2023direct}. Smooth temporal tests additionally control the weighting of velocity information within a window. Our estimator applies this foundation to the speed-coordinate action of an input-conditioned generator, with a calibrated car-following field as the nominal response. Common-dictionary controls distinguish the contribution of temporal testing from the representation and physical prior. Relative to our earlier study of finite-difference, logarithm, and resolvent estimators for autonomous synthetic ACC dynamics~\cite{meng2024koopman}, the present formulation retains measured leader motion as an explicit input and estimates a compact physical correction on finite observation windows.

\subsection{Response-based behavioral characterization}
Vehicle-response models describe how an automated vehicle or human driver reacts to surrounding traffic, supplying information for predictive interaction and safety analysis~\cite{mahbub2023safety,li2025robotic}. Beyond vehicle dynamics modeling and motion prediction, model-fitting residuals can characterize how consistently observed behavior follows an identified response law.

PI-Koopman connects conditional prediction and behavioral characterization through a physically anchored acceleration-response field. Its fitting residuals quantify the agreement between observed vehicle motion and the identified dynamics after accounting for spacing, follower speed, and measured leader motion. Smaller residuals indicate more consistent, model-explainable responses, as expected under active ACC, whereas larger residuals indicate greater deviations that may be associated with human driving~\cite{li2023automated}.

\begin{figure*}[!t]
\centering
\includegraphics[width=\textwidth]{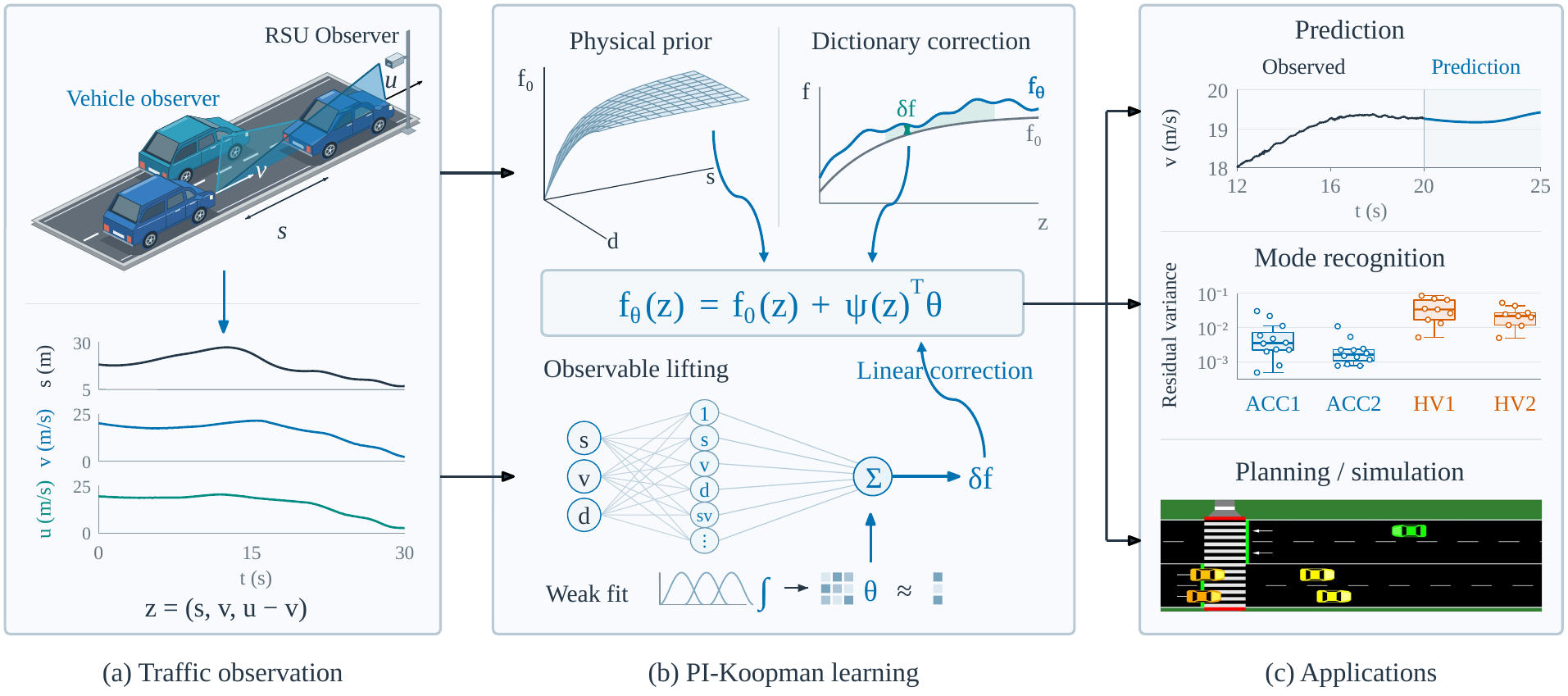}
\caption{PI-Koopman framework from traffic observation, physics-informed Koopman learning, and downstream applications.}
\label{fig:workflow}
\end{figure*}

\section{Method}
\label{sec:method}
PI-Koopman identifies a continuous-time car-following response from a short encounter. A calibrated physical law provides the nominal acceleration, while a weak-form estimator learns a dictionary-based correction from velocity observations. The fitted response supports conditional prediction and residual-based following-mode recognition (Fig.~\ref{fig:workflow}).

\subsection{Input-conditioned response representation}
Let $s$ denote longitudinal gap, $v$ follower speed, $u$ leader speed projected onto the follower's longitudinal axis, and $d=u-v$ relative speed. With $z=(s,v,d)^\top$, the following dynamics are
\begin{equation}
\dot s=u-v+c(t),\qquad \dot v=f(z)+\rho(t),
\label{eq:dynamics}
\end{equation}
where $f$ is the acceleration response, $c(t)$ accounts for observation-frame rotation, and $\rho(t)$ represents unresolved acceleration. Prediction assumes a straight road ($c=0$).

For the modeled dynamics with $c=\rho=0$, let $x=(s,v)^\top$ and $F(x,u)=(u-v,\,f(s,v,u-v))^\top$. The Koopman generator describes the instantaneous evolution of an observable under a vector field~\cite{mauroy2019koopman}. Its input-conditioned action on the follower-speed observable $p_v(x)=v$ is
\begin{equation}
\mathcal L_u p_v(x)=\nabla_x p_v(x)^\top F(x,u)=f(z).
\label{eq:vectorfield}
\end{equation}

PI-Koopman represents this generator action as a calibrated physical response $f_0$ plus a finite-dimensional correction
\begin{equation}
f_\theta(z)=f_0(z)+\psi(z)^\top\theta,
\label{eq:library}
\end{equation}
where $\psi(z)\in\mathbb R^p$ is the dictionary of state-input features and $\theta\in\mathbb R^p$ contains encounter-specific coefficients.

\subsection{Physical priors and target-specific correction}
Constant-time-headway (CTH) model~\cite{swaroop1994comparision} and the intelligent driver model (IDM)~\cite{treiber2000congested} provide two alternative physical priors. Their acceleration laws are
\begin{equation}
f_{\rm CTH}(z)=\alpha(s-\eta-\tau v)+\beta d,
\label{eq:cth}
\end{equation}
\begin{align}
f_{\rm IDM}(z)&=a_{\max}\!\left[1-\left(\frac{v}{v_{\rm des}}\right)^4-\left(\frac{s^*(z)}{s}\right)^2\right],\nonumber\\
s^*(z)&=s_0+\max\!\left(0,\tau v-\frac{vd}{2\sqrt{a_{\max}b_{\rm c}}}\right).
\label{eq:idm}
\end{align}

In CTH, $\alpha$ and $\beta$ are feedback gains, $\eta$ is standstill spacing, and $\tau$ is desired time headway. In IDM, $a_{\max}$ is the acceleration parameter, $b_{\rm c}$ comfortable deceleration, $v_{\rm des}$ desired speed, $s_0$ standstill spacing, and $s^*(z)$ desired spacing. Its closing-speed term uses $-d=v-u$.

Let $\vartheta$ collect the parameters of either prior. At observation times $t_i$, the simulated states $\widehat s_\vartheta(t_i)$ and $\widehat v_\vartheta(t_i)$ are obtained from the initial observed follower state under the measured leader speed. We calibrate $\vartheta$ by bounded trajectory output-error minimization
\begin{equation}
\begin{aligned}
J(\vartheta)=\sum_i\Bigg[&\left(\frac{\widehat s_\vartheta(t_i)-s(t_i)}{10\,\mathrm m}\right)^2\\
&+\left(\frac{\widehat v_\vartheta(t_i)-v(t_i)}{1\,\mathrm{m/s}}\right)^2\Bigg].
\end{aligned}
\label{eq:physicalfit}
\end{equation}

The calibrated parameters are fixed during correction fitting.

PI-Koopman-CTH uses the quadratic dictionary
\begin{equation}
\phi_{\rm CTH}=(1,s,v,d,s^2,v^2,d^2,sv,sd,vd)^\top,
\end{equation}
while PI-Koopman-IDM uses
\begin{equation}
\phi_{\rm IDM}=\big(1,s,v,d,e_0^2,e_0d,d^2,\tanh(d/v_{\rm sc}),e_0|e_0|\big)^\top,
\end{equation}
where $e_0=[s-3\,\mathrm m-(1.5\,\mathrm s)v]/(10\,\mathrm m)$ and $v_{\rm sc}=1\,\mathrm{m/s}$. For either raw dictionary $\phi(z)\in\mathbb R^p$ ($p=10$ for CTH and $p=9$ for IDM), the standardized features are
\begin{equation}
\psi_0=1,\qquad \psi_k=\frac{\phi_k-\mu_k}{\sigma_k},\quad k=1,\ldots,p-1,
\label{eq:scaling}
\end{equation}
where $\mu_k$ and $\sigma_k$ are the target-history mean and population standard deviation of feature $\phi_k$. Scales at or below $10^{-8}$ are replaced by one to avoid division by a near-zero value.

\subsection{Weak estimation from velocity observations}
We estimate the correction through weak temporal equations rather than pointwise acceleration estimates, following weak-form system identification~\cite{messenger2021weak}. Define the remaining acceleration discrepancy as $\epsilon_\theta(t)=f(z(t))-f_\theta(z(t))+\rho(t)$. For a differentiable test function $w$ on a history interval $[a,b]$, integration by parts gives $b_w=A_w\theta+\int_a^b w(t)\epsilon_\theta(t)\,\mathrm dt$, where the regression row $A_w$ and scalar response $b_w$ are
\begin{align}
(A_w)_k&=\int_a^b w(t)\psi_k(z(t))\,\mathrm dt,\nonumber\\
b_w&=[wv]_a^b-\int_a^b w'(t)v(t)\,\mathrm dt
-\int_a^b w(t)f_0(z(t))\,\mathrm dt.
\label{eq:weak}
\end{align}

For support duration $T=b-a$ and normalized time $q=(t-a)/T$, we use the compact test family
\begin{equation}
w_\kappa(t)=\frac{e^{-\kappa q}B(q)}{T I_\kappa},\qquad
I_\kappa=\int_0^1 e^{-\kappa q}B(q)\,\mathrm dq,
\label{eq:kernels}
\end{equation}
with $B(q)=16q^2(1-q)^2$. The test vanishes at both endpoints. Setting $\kappa=0$ gives a symmetric compact test, and $\kappa>0$ emphasizes earlier observations within the support. Both PI-Koopman branches use $\kappa=1$.

For samples $t_0=a<\cdots<t_N=b$, let $v_i=v(t_i)$ and $q_i=(t_i-a)/T$. Composite trapezoidal quadrature, denoted $Q_h$ with nominal sampling interval $h$, is
\begin{equation}
Q_h(g)=\sum_{i=0}^{N-1}\frac{t_{i+1}-t_i}{2}(g_i+g_{i+1}).
\label{eq:quadrature}
\end{equation}

We normalize discrete weights as $w_i=K_i/Q_h(K)$, where $K_i=e^{-\kappa q_i}B(q_i)$. The resulting summation-by-parts equations are
\begin{align}
\widehat y&=\sum_{i=0}^{N-1}\frac{w_i+w_{i+1}}{2}(v_{i+1}-v_i),
\label{eq:discrete}\\
\widehat A_k&=Q_h(w\psi_k),\qquad
\widetilde y=\widehat y-Q_h(wf_0).
\label{eq:hybrid}
\end{align}

The discretization is exact for constant acceleration. Each complete support with at least three samples yields one regression row, and successive supports overlap by approximately half their duration. Stacking $n_{\rm fit}$ supports into a design matrix $\widehat A$ and response vector $\widetilde y$, we estimate
\begin{equation}
\widehat\theta=\arg\min_\theta\left\{
\frac{\|\widehat A\theta-\widetilde y\|_2^2}{2n_{\rm fit}}
+\frac{\gamma}{2}\|\theta\|_2^2\right\},
\label{eq:ridge}
\end{equation}
where $\gamma\geq0$ is the ridge coefficient. Prediction selects the support duration and $\gamma$ as described in Section~\ref{sec:predictionprotocol}, while recognition uses the fixed configuration in Section~\ref{sec:recognitionprotocol}.

For conditional prediction, the fitted field is propagated from the observed follower state with $\dot s=u-v$ and $\dot v=f_{\widehat\theta}(s,v,u-v)$. Fourth-order Runge-Kutta integration uses a maximum step of $0.04$ s and piecewise-linear interpolation of the supplied leader speed $u(t)$. In the controlled field-recovery experiments, $f_0=0$ and a common dictionary is used to compare finite differences (FD), ordinary integral regression ($w=1/T$), compact weak testing ($\kappa=0$), and exponential weak testing ($\kappa>0$). FD uses adjacent velocity increments, with endpoint-averaged features in the paired-rate experiment.

\subsection{Response consistency from identification residuals}
\label{sec:residualmethod}
After fitting the complete field $\widehat f=f_0+\psi^\top\widehat\theta$ to an encounter, we evaluate velocity-integral residuals on intervals $[t_j,t_j+T_{\rm res}]$, where $t_j$ is the start time and $T_{\rm res}$ the evaluation duration
\begin{equation}
r_j=\frac{v(t_j+T_{\rm res})-v(t_j)-\int_{t_j}^{t_j+T_{\rm res}}\widehat f(z(t))\,\mathrm dt}{T_{\rm res}}.
\label{eq:response-residual}
\end{equation}

Each residual is computed along observed states using ordinary integration. Across $n_{\rm res}$ supports, we measure within-encounter response consistency by the sample variance
\begin{equation}
S_r=\frac{1}{n_{\rm res}-1}\sum_{j=1}^{n_{\rm res}}(r_j-\bar r)^2,
\qquad \bar r=\frac{1}{n_{\rm res}}\sum_{j=1}^{n_{\rm res}}r_j.
\label{eq:residual-variance}
\end{equation}

Lower $S_r$ indicates smaller residual fluctuations around the fitted response. Residual RMS additionally reflects any systematic bias. Each encounter is fitted without its driving-mode label. A threshold $\lambda_r$, selected to maximize training balanced accuracy, assigns the predicted following mode as
\begin{equation}
\widehat m=\begin{cases}
\mathrm{ACC},&S_r\leq\lambda_r,\\
\mathrm{Manual},&S_r>\lambda_r.
\end{cases}
\label{eq:mode-decision}
\end{equation}

The decision is evaluated on held-out runs (Section~\ref{sec:recognitionprotocol}).

\begin{figure*}[!t]
\centering\includegraphics[width=\textwidth]{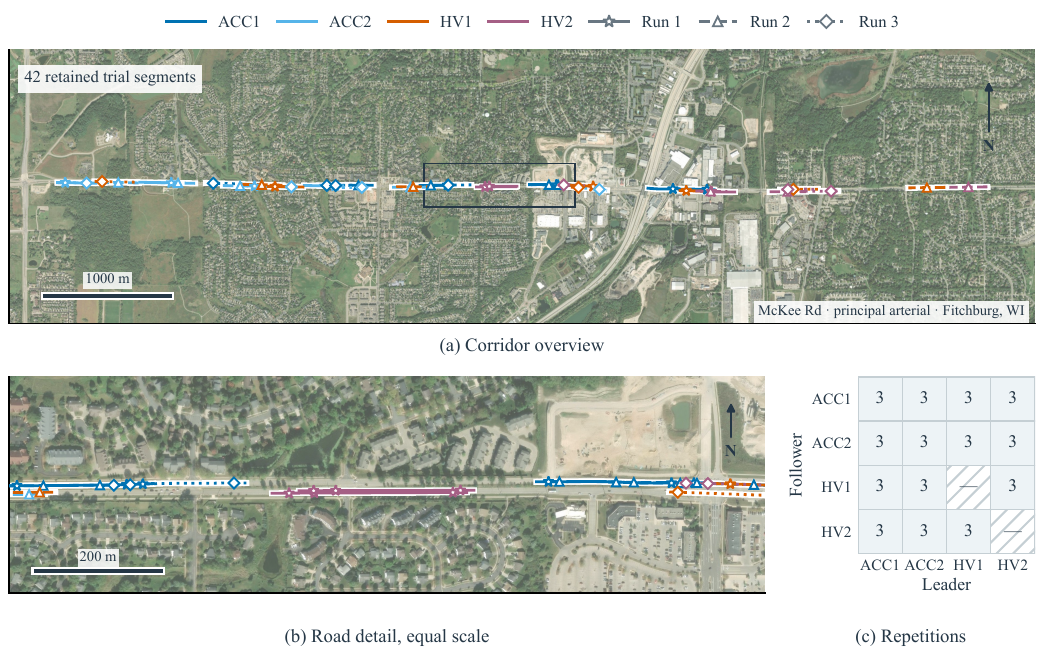}
\caption{McKee Rd collection corridor and repeated leader-follower design. Satellite imagery from Esri, Vantor, Earthstar Geographics, and the GIS User Community.}
\label{fig:collectiondesign}
\end{figure*}

\begin{figure*}[!t]
\centering\includegraphics[width=\textwidth]{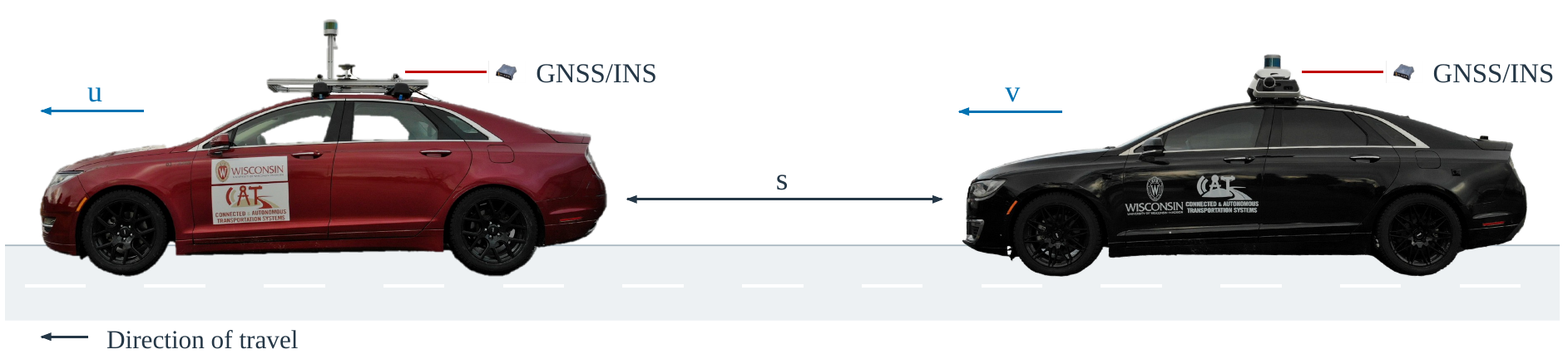}
\caption{CATS-CF vehicle platforms with GNSS/INS.}
\label{fig:vehiclesetup}
\end{figure*}

\section{Experiment}
\label{sec:experiment}
The experiments connect short-encounter prediction, noise-robust response recovery, and following-mode recognition. OpenACC evaluates identification across production vehicles, while CATS-CF pairs repeated ACC settings and drivers with synchronized high-rate measurements. Together, they test predictive accuracy and the response consistency that supports downstream use. 

\subsection{OpenACC production-vehicle benchmark}
OpenACC provides commercial ACC trajectories across test-track and public-road operation~\cite{makridis2021openacc}. The selected AstaZero and Ispra-Casale Monferrato campaigns cover five follower platforms. AstaZero used OxTS RT-Range S differential GNSS/INS, acquiring above 100 Hz and releasing observations at 10 Hz. Casale used Ublox 9 and on-board diagnostics, with cubic-spline interpolation to the released 10-Hz grid. The analysis uses the published speeds, bumper spacing, and following-mode labels.

Each follower sequence within a source file is partitioned into nonoverlapping 30-s blocks with finite, nonnegative speeds, positive spacing, consistent 0.1-s timestamps, and an unchanged mode label. Up to six eligible blocks per sequence are selected at evenly spaced chronological indices. The short-encounter prediction benchmark uses the resulting 236 ACC episodes from 17 source files, comprising 192 AstaZero and 44 Casale episodes.

\subsection{CATS-CF repeated-vehicle dataset}
CATS-CF was collected on McKee Rd, a principal arterial in Fitchburg, WI. The repeated design in Fig.~\ref{fig:collectiondesign} holds the vehicle platform fixed while varying the following mode and leader. Two Lincoln MKZ vehicles carry synchronized Bynav GNSS/INS systems (Fig.~\ref{fig:vehiclesetup}). ACC1 and ACC2 denote two following-distance settings, and HV1 and HV2 denote two human drivers. Each ACC follower has four leader labels, while each human follower follows both ACC settings and the other driver. Three repetitions of these 14 conditions give 42 trials. A 30-s record per trial supports prediction, and nominal 15-s records support paired sensing and recognition.

Both vehicles retain 125-Hz navigation outputs, attitude, inertial channels, and positioning-quality metadata. RTK-fixed solutions account for 99.53\% of vehicle samples in the 15-s records and 96.79\% in the 30-s records. Synchronized continuous channels are linearly interpolated onto a common 125-Hz grid. Positions are transformed into a per-trial local frame and projected along the follower heading, with a 3-m sensor-spacing correction for bumper gap. Velocity vectors projected onto the same axis give follower and leader speeds. In the 15-s records, integrating relative speed reproduces measured gap with a mean trial RMS discrepancy of 0.0211 m. The released data retain local coordinates and relative time after removal of absolute geographic coordinates and timestamps.

This measurement design supplies paired views of the same motion at different observation rates. Its repeated settings and drivers also connect an identified response to a known following mode, providing a common basis for the sensing and recognition experiments.

\subsection{Matched prediction and evaluation}
\label{sec:predictionprotocol}
Every model identifies a newly observed ACC follower from the same target history and then predicts its motion for 5 s. Forecasts start from the measured follower state and use the measured future leader trajectory, with the recorded follower heading defining the CATS-CF longitudinal frame. The follower states subsequently propagate freely. This evaluates a conditional response model; deployment additionally requires a forecast or planned trajectory for the leader.

The duration comparison uses nested histories $[20-B,20]$ s, with $B=5,7,11$ s, and the common future $(20,25]$ s. Both datasets use 10-Hz observations, giving 51, 71, or 111 endpoint-inclusive samples. OpenACC retains its released grid, and CATS-CF interpolates its 125-Hz record onto exact 0.1-s timestamps. The rate comparison fixes a 7-s history and uses CATS-CF observations at 1, 5, 10, 25, and 125 Hz. Three origins $o\in\{15,20,24\}$ s define histories $[o-7,o]$ and futures $(o,o+5]$ s. Their errors are averaged within each trial. All models share these windows, and exact-divisor rates retain samples from the same 125-Hz parent.

CTH and IDM provide physical baselines, and PI-Koopman-CTH and PI-Koopman-IDM add their respective generator corrections. The neural comparators are an acceleration DNN, an LSTM, and a projected Deep Koopman model adapted from input-driven lifted prediction~\cite{tian2025physically}. Parameters and normalization are fitted separately to each target history. Configuration selection excludes the target OpenACC source file or CATS-CF run index across all conditions. For each duration-rate-origin condition, the selection score is speed RMSE plus gap RMSE divided by 10.

Physical calibration uses two starts with at most 400 function evaluations each. PI support candidates are $T=\max(2/f_s,T_0)$ for $T_0\in\{0.5,1,2\}$ s at sampling rate $f_s$, with ridge values $10^{-3},0.1,10$ and $\kappa=1$. The DNN has two tanh layers of width 16 or 32, the LSTM has 8 or 32 hidden units, and Deep Koopman adds six latent coordinates. Neural selection varies rollout length and optimizer checkpoints up to 6000 updates. Compact and Deep Koopman results average two initializations per encounter, and expanded DNN/LSTM results average five. The LSTM uses a 1-s observed warmup during training and forecasting.

Physical and PI fields use continuous-time integration. Neural selection includes continuous-integration DNN and Deep Koopman candidates, while recurrent and acquisition-step candidates update on the observation clock. All models use the same native-grid future leader input, interpolated at integration substeps. Forecasts are scored at 10 Hz for OpenACC and 125 Hz for CATS-CF.

For $q\in\{v,s\}$, the prediction metric is
\begin{equation}
 \mathrm{RMSE}_q=\left\{\frac{1}{N}\sum_{j=1}^{N}[\widehat q(t_j)-q(t_j)]^2\right\}^{1/2},
\end{equation}
excluding the common initial state. Tables pair speed and gap RMSE in m/s and m, weighting the 236 OpenACC or 24 CATS-CF ACC encounters equally. The duration-table Mean gives each history length equal weight. Parentheses beside the PI-Koopman Mean values give $100(1-\mathrm{Mean}_{\mathrm{PI}}/\mathrm{Mean}_{\mathrm{prior}})$ relative to the respective CTH or IDM prior, with positive percentages indicating lower error. Paired bootstrap intervals resample entire OpenACC files or CATS-CF leader-follower conditions, retaining all matched observations within each of the 17 files or eight ACC conditions.

\subsection{Temporal estimation under controlled sensing}
\label{sec:pairedtasks}
The temporal-estimation experiments compare finite differences, ordinary integration, compact weak testing, and exponential weak testing with the same ten-term quadratic dictionary and $f_0=0$. Holding representation and observations fixed isolates how the generator response is estimated.

An exact acceleration law supplies a field-reconstruction target. With $e=s-4-1.3v$,
\begin{equation}
 f(z)=0.12e+0.55d+0.002e^2-0.008d^2.
 \label{eq:synthetic}
\end{equation}

An affine control removes the quadratic terms. Eighteen 20-s training trajectories and twelve independent test trajectories use randomized initial states and sinusoidal leader excitation. Leader speed combines a base in $[14,24]$ m/s with amplitudes 1.5 and 0.5 m/s at angular frequencies 0.45 and 1.2 rad/s. Fourth-order Runge-Kutta integration uses 0.008-s steps. Sampling at 1/5/25 Hz is crossed with independent Gaussian speed errors of standard deviation 0/0.05/0.15 m/s and twice these values for gap. Eight realizations perturb gap and both speeds, with relative speed reconstructed from the perturbed channels. Estimators share ridge $10^{-6}$ with an unpenalized intercept. Integrated estimators use 4-s supports, and exponential tests use $\kappa=3$. Acceleration RMSE is evaluated against the exact law on independent clean trajectories.

For paired real-data interventions, all 42 CATS-CF trials use $[2,10]$ s for calibration and the same 41 uncorrupted 5-Hz states as field probes. The 125/25/5/1-Hz fits contain 1001/201/41/9 observations. Each fit standardizes its observed features and uses a mean-squared regression loss with ridge 0.1 on all coefficients. Finite differences use adjacent velocity increments and endpoint-averaged features. Integrated methods use seven overlapping 2-s supports beginning one second apart, normalized trapezoidal kernels, and summation-by-parts responses, with $\kappa=1$ for exponential weighting.

Independent and temporally correlated Gaussian errors have standard deviations of 0.5 m in gap and 0.1 m/s in each speed, with a 0.3-s correlation time in the latter case. Three realizations perturb the 125-Hz parent before subsampling, preserving shared-timestamp errors. For method $m$, rate $r$, realization $\epsilon$, and common probes $z_j$, define
\begin{align}
 D_{\mathrm{rate}}&=\operatorname{RMS}_j\big[f_{m,r,\epsilon}(z_j)-f_{m,125,\epsilon}(z_j)\big],\\
 D_{\mathrm{ref}}&=\operatorname{RMS}_j\big[f_{m,r,\epsilon}(z_j)-f_{m,125,0}(z_j)\big].
\end{align}

$D_{\mathrm{rate}}$ measures agreement across acquisition rates within one measurement realization. $D_{\mathrm{ref}}$ measures perturbation from each estimator's clean high-rate fit, isolating corruption sensitivity at 125 Hz. A matched synthetic control uses 12 parameterized quadratic laws with three leader/initial-state realizations each to link consistency to exact-law accuracy. Runs and corruptions are averaged within each of the 14 recorded conditions or 12 synthetic objects. Paired 95\% intervals use 2000 group-bootstrap resamples.

\subsection{Following-mode recognition from response consistency}
\label{sec:recognitionprotocol}
The downstream experiment uses identification residuals to distinguish ACC from manual following. All 42 nominal 15-s CATS-CF encounters are used at 25 Hz, with 24 ACC and 18 manual records. Each encounter receives an independent CTH, IDM, PI-Koopman-CTH, and PI-Koopman-IDM fit using its 375 observations. PI fitting uses the dictionaries in Section~\ref{sec:method}, 1-s exponential weak supports, $\kappa=1$, and ridge 0.1. Mode labels enter the subsequent threshold rule, while response fitting uses the observed trajectories.

The fitted response is evaluated on thirty overlapping 1-s residual supports within the same encounter, each containing 26 samples with a 12-sample stride. Trapezoidal integration gives~\eqref{eq:response-residual}, and residual variance $S_r$ supplies the recognition feature. A lower value indicates more consistent agreement with one local response and is assigned to ACC.

Three folds each hold out one complete run index, comprising eight ACC and six manual encounters. The other 28 trials select the threshold that maximizes balanced accuracy (BA). Thus settings and drivers recur across repetitions, while the recognition decision is tested on a held-out run. Pooled test predictions give BA and class recalls, and AUC ranks encounters by negative residual variance. A 24-trial subset with common ACC leader labels, containing 12 ACC and 12 manual followers, repeats threshold selection to evaluate separation under shared leader contexts.

\begin{table*}[!t]
\centering\footnotesize
\caption{ACC prediction across observation durations.}
\label{tab:avlearning}
\setlength{\tabcolsep}{3pt}
\begin{tabular}{llcccc}\toprule
\multicolumn{6}{c}{Speed / gap RMSE (m/s / m), $\downarrow$ Lower is better\qquad Mean reduction (\%), $\uparrow$ Higher is better}\\\midrule
Dataset & Method & 5 s & 7 s & 11 s & Mean\\\midrule
\multirow[c]{7}{*}[-1.5ex]{\shortstack[l]{OpenACC\\(236 encounters)}} & CTH & \textbf{0.331} / 0.626 & 0.333 / 0.653 & 0.370 / 0.766 & 0.345 / 0.681\\
 & IDM & 0.347 / 0.657 & 0.352 / 0.663 & 0.377 / 0.765 & 0.359 / 0.695\\
\cmidrule(lr){2-6}
 & \textbf{PI-Koopman-CTH (ours)} & 0.333 / \textbf{0.621} & \textbf{0.327} / \textbf{0.634} & \textbf{0.362} / 0.741 & \textbf{0.341}\,(+1.1\%) / \textbf{0.665}\,(+2.4\%)\\
 & \textbf{PI-Koopman-IDM (ours)} & 0.382 / 0.690 & 0.400 / 0.761 & 0.378 / 0.734 & 0.386\,(-7.7\%) / 0.728\,(-4.8\%)\\
\cmidrule(lr){2-6}
 & DNN & 0.435 / 0.706 & 0.415 / 0.688 & 0.387 / 0.672 & 0.412 / 0.689\\
 & LSTM & 0.418 / 0.680 & 0.400 / 0.659 & 0.399 / \textbf{0.658} & 0.406 / 0.666\\
 & Deep Koopman & 0.880 / 1.271 & 0.654 / 1.055 & 0.629 / 0.992 & 0.721 / 1.106\\
\midrule
\multirow[c]{7}{*}[-1.5ex]{\shortstack[l]{CATS-CF\\(24 encounters)}} & CTH & 0.428 / 0.771 & 0.445 / 0.913 & 0.493 / 1.019 & 0.455 / 0.901\\
 & IDM & 0.519 / 0.921 & 0.500 / 0.963 & 0.373 / 0.835 & 0.464 / 0.906\\
\cmidrule(lr){2-6}
 & \textbf{PI-Koopman-CTH (ours)} & \textbf{0.420} / \textbf{0.744} & \textbf{0.423} / 0.868 & 0.479 / 0.989 & \textbf{0.441}\,(+3.2\%) / 0.867\,(+3.8\%)\\
 & \textbf{PI-Koopman-IDM (ours)} & 0.542 / 0.913 & 0.465 / 0.892 & \textbf{0.337} / \textbf{0.676} & 0.448\,(+3.5\%) / \textbf{0.827}\,(+8.7\%)\\
\cmidrule(lr){2-6}
 & DNN & 0.502 / 0.776 & 0.618 / 1.054 & 0.458 / 0.823 & 0.526 / 0.884\\
 & LSTM & 0.554 / 0.919 & 0.511 / \textbf{0.813} & 0.497 / 0.776 & 0.521 / 0.836\\
 & Deep Koopman & 0.607 / 0.859 & 0.566 / 0.880 & 0.556 / 0.824 & 0.577 / 0.854\\
\bottomrule\end{tabular}
\end{table*}

\begin{figure*}[!t]
\centering\includegraphics[width=\textwidth]{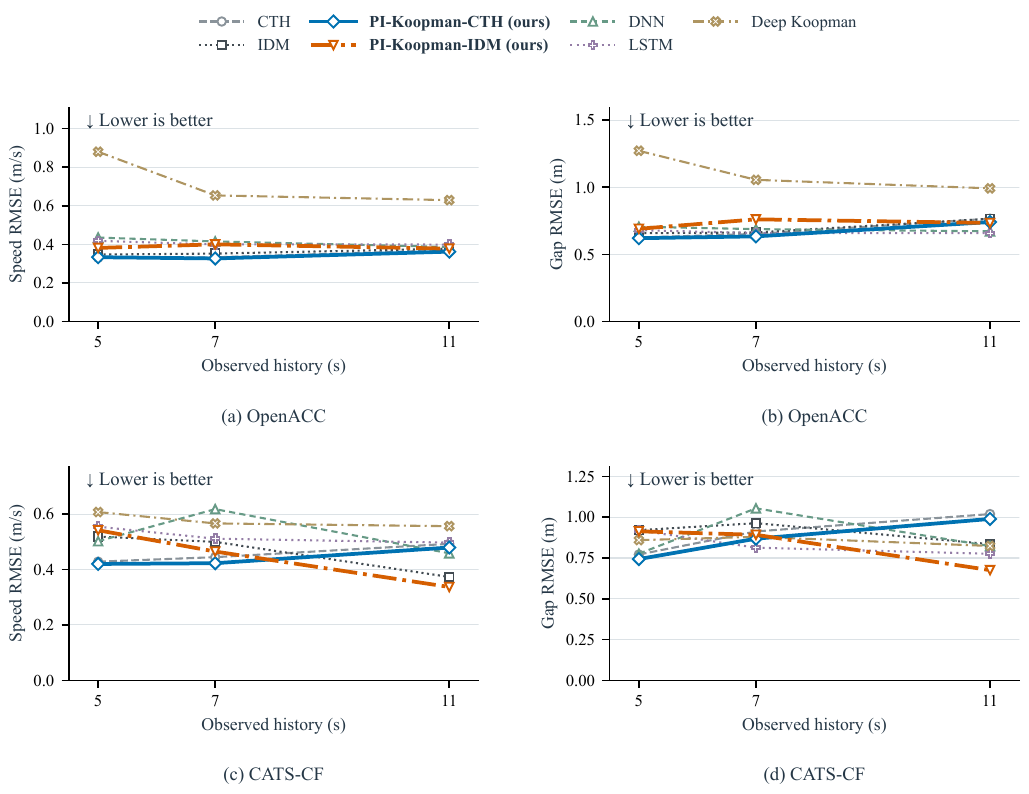}
\caption{Prediction error versus observation duration.}
\label{fig:avbudget}
\end{figure*}

\begin{figure*}[!t]
\centering\includegraphics[width=\textwidth]{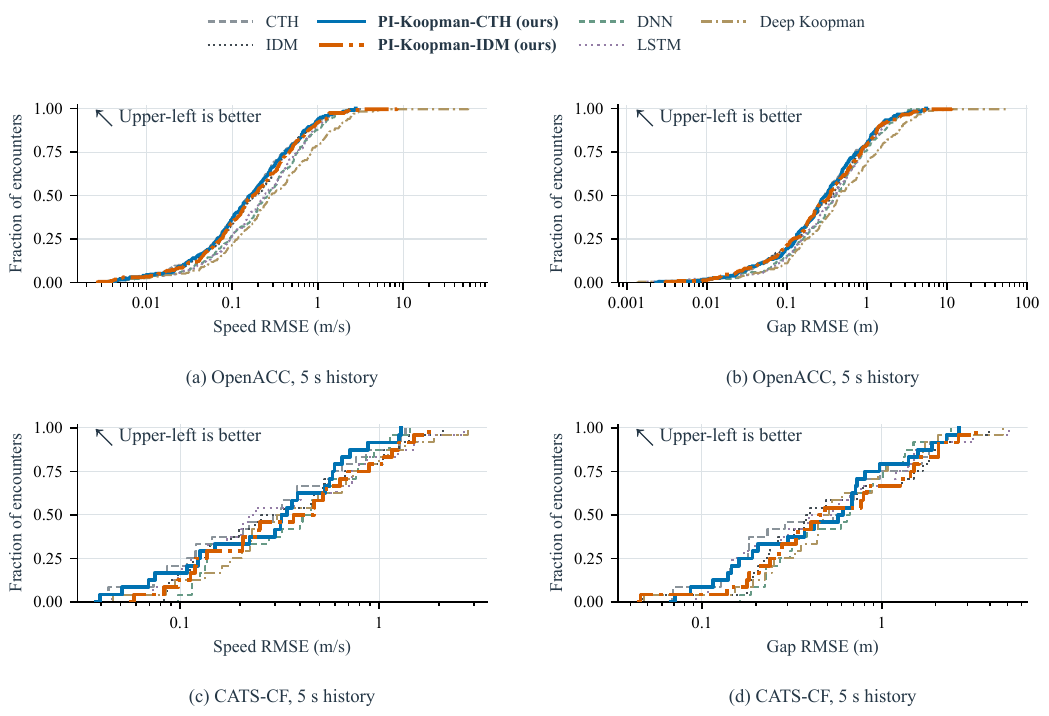}
\caption{Encounter-level prediction errors with 5-s histories.}
\label{fig:encountercdf}
\end{figure*}

\section{Results}
\subsection{Data-efficient prediction from short encounters}
\label{sec:prediction}

PI-Koopman combines short-history adaptation with accurate conditional prediction across both vehicle collections. With 5-s histories, PI-Koopman-CTH reduces mean speed RMSE by 20.3\% on OpenACC and 16.4\% on CATS-CF, relative to LSTM and DNN, the respective best neural speed comparators (Table~\ref{tab:avlearning}). The paired 95\% cluster-bootstrap intervals for speed reduction are [11.0, 31.9]\% and [$-18.0$, 39.6]\%, respectively. Physical structure and a compact correction provide an effective representation when each target contributes a few seconds of observations.

The encounter-level distributions in Fig.~\ref{fig:encountercdf} complement these dataset means. With 5-s histories, PI-Koopman-CTH has lower speed error than LSTM in 61.0\% of OpenACC encounters, with median errors of 0.162 and 0.237 m/s, respectively. Each curve gives the fraction of encounters below an error threshold, so movement toward the upper left indicates better prediction.

PI-Koopman achieves the lowest error in nine of the twelve comparisons across datasets, durations, and prediction metrics (Fig.~\ref{fig:avbudget}). Averaged equally over 5-, 7-, and 11-s histories, PI-Koopman-CTH achieves the lowest mean speed errors on OpenACC and CATS-CF, 0.341 and 0.441 m/s, respectively. The respective speed reductions are 16.0\% and 15.4\% relative to the best neural means, and 1.1\% and 3.2\% relative to the best calibrated physical means. The CTH and IDM branches achieve the lowest mean gap errors on OpenACC and CATS-CF, 0.665 and 0.827 m, improving their physical priors by 2.4\% and 8.7\%, respectively. These gains reflect target-data efficiency under the common encounter-level fitting protocol.

\begin{table*}[!t]
\centering\footnotesize
\caption{ACC prediction across acquisition rates (7-s history).}
\label{tab:protocol-rate-speed}
\setlength{\tabcolsep}{6pt}
\begin{tabular}{lccccc}\toprule
\multicolumn{6}{c}{Speed / gap RMSE (m/s / m), $\downarrow$ Lower is better}\\\midrule
Method & 1 Hz & 5 Hz & 10 Hz & 25 Hz & 125 Hz\\\midrule
CTH & 0.347 / 0.681 & 0.365 / 0.721 & 0.369 / 0.729 & 0.373 / 0.739 & 0.375 / 0.743\\
IDM & 0.432 / 0.853 & 0.445 / 0.891 & 0.448 / 0.898 & 0.466 / 0.932 & 0.468 / 0.937\\
\midrule
\textbf{PI-Koopman-CTH (ours)} & \textbf{0.342} / \textbf{0.674} & \textbf{0.352} / \textbf{0.694} & \textbf{0.357} / \textbf{0.705} & \textbf{0.357} / \textbf{0.700} & \textbf{0.360} / \textbf{0.707}\\
\textbf{PI-Koopman-IDM (ours)} & 0.429 / 0.845 & 0.424 / 0.821 & 0.418 / 0.826 & 0.426 / 0.842 & 0.429 / 0.849\\
\midrule
DNN & 0.461 / 0.799 & 0.496 / 0.870 & 0.489 / 0.857 & 0.501 / 0.878 & 0.488 / 0.863\\
LSTM & 0.510 / 0.893 & 0.457 / 0.795 & 0.468 / 0.822 & 0.500 / 0.903 & 0.504 / 0.917\\
Deep Koopman & 0.555 / 0.919 & 0.577 / 0.912 & 0.535 / 0.840 & 0.654 / 0.982 & 0.697 / 1.035\\
\bottomrule\end{tabular}
\end{table*}

\begin{figure*}[tp]
\centering\includegraphics[width=.94\textwidth]{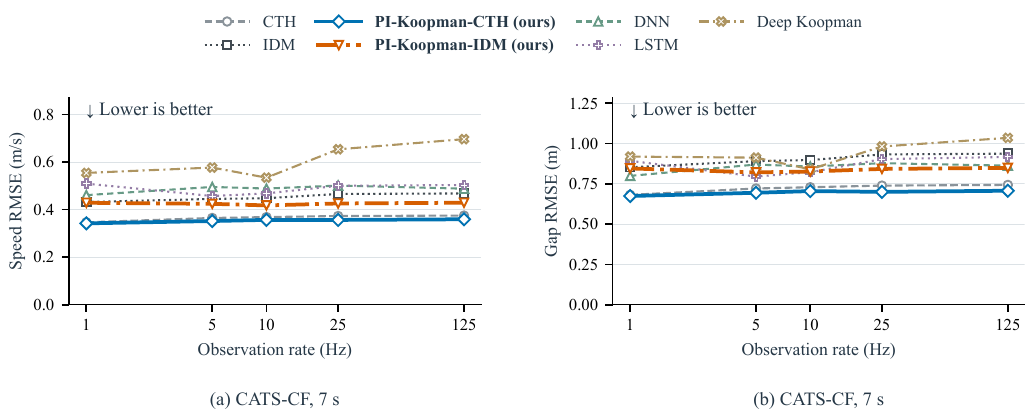}
\caption{Prediction error versus observation rate with 7-s histories.}
\label{fig:acquisitionbudget}
\end{figure*}

\subsection{Prediction across observation rates}
\label{sec:windowrate}

The retained high-rate CATS-CF records support prediction across acquisition clocks using paired observations of the same encounters. With a 7-s history and equal averaging over the 15-, 20-, and 24-s prediction origins, PI-Koopman-CTH has the lowest mean speed and gap errors among all seven models at every tested rate from 1 to 125 Hz (Table~\ref{tab:protocol-rate-speed}). Its reductions relative to the lowest neural mean are 23.0-28.7\% for speed and 12.7-20.3\% for gap.

Both PI corrections improve on their own physical priors in speed and gap at all five rates. At 10 Hz, PI-Koopman-CTH gives 0.357 m/s / 0.705 m, reducing CTH's errors by 3.2\%/3.3\%. The IDM correction gives reductions of 6.8\%/7.9\%. Across the full rate sweep, PI-Koopman-CTH has speed errors of 0.342-0.360 m/s and gap errors of 0.674-0.707 m.

Figure~\ref{fig:acquisitionbudget} shows how the continuous-time response maintains predictive accuracy across observation clocks. The following common-dictionary experiments examine the temporal-estimation mechanism used in estimating response fields.

\begin{table*}[!t]
\centering\footnotesize
\caption{Known-law field RMSE at 25 Hz ($10^{-3}$ m/s$^2$; $\downarrow$ lower is better).}
\label{tab:evidenceknownlaw}
\setlength{\tabcolsep}{4pt}
\begin{tabular}{llrrrr}\toprule
Law & Noise SD (m/s) & Finite difference & Ordinary integral & \textbf{Weak (ours)} & \textbf{Exp. weak (ours)}\\\midrule
\multirow[c]{3}{*}{Affine} & 0.00 & 0.017 & \textbf{0.010} & 0.034 & 0.032\\
 & 0.05 & 5.871 $\pm$ 1.704 & 5.379 $\pm$ 1.108 & \textbf{2.479} $\pm$ 0.495 & 2.582 $\pm$ 0.351\\
 & 0.15 & 27.415 $\pm$ 7.662 & 16.103 $\pm$ 3.227 & \textbf{7.424} $\pm$ 1.516 & 7.744 $\pm$ 1.100\\
\midrule
\multirow[c]{3}{*}{Quadratic} & 0.00 & 0.022 & \textbf{0.011} & 0.037 & 0.039\\
 & 0.05 & 6.449 $\pm$ 2.177 & 5.833 $\pm$ 1.610 & 3.083 $\pm$ 1.546 & \textbf{3.022} $\pm$ 1.333\\
 & 0.15 & 35.513 $\pm$ 12.666 & 17.234 $\pm$ 4.815 & 9.212 $\pm$ 4.444 & \textbf{9.055} $\pm$ 3.761\\
\bottomrule\end{tabular}

\end{table*}

\begin{figure*}[!t]
\centering\includegraphics[width=.94\textwidth]{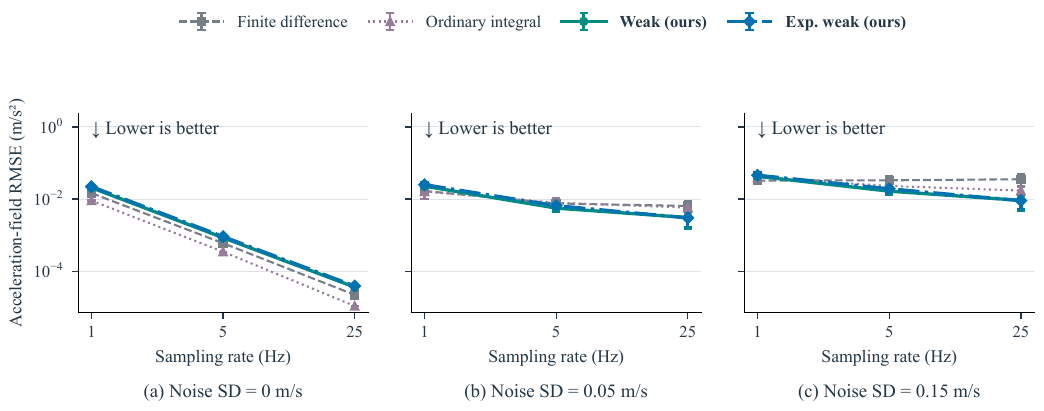}
\caption{Acceleration-field recovery across sampling rates and noise levels.}
\label{fig:field}
\end{figure*}

\subsection{Reconstruction of noisy response fields}
\label{sec:fieldresults}

Weak temporal testing improves field recovery by combining velocity observations over a physical interval. For the quadratic generating law at 25 Hz and speed-noise standard deviation 0.15 m/s, exponential weak estimation gives acceleration RMSE 0.00905 m/s$^2$, compared with 0.03551 for finite differences and 0.01723 for ordinary integration (Table~\ref{tab:evidenceknownlaw}). The respective reductions are 74.5\% and 47.5\%, with compact weak estimation giving 0.00921 m/s$^2$. With the dictionary and observations fixed, the improvement over ordinary integration identifies the benefit of smooth temporal weighting.

The same comparison spans affine and quadratic generating laws, three noise levels, and three observation rates. Table~\ref{tab:evidenceknownlaw} reports mean and standard deviation over eight error realizations at 25 Hz, and Fig.~\ref{fig:field} shows the rate sweep with one-std error bars. Denser observations refine quadrature and provide more velocity samples within the fixed support. 

\begin{table}[!b]
\centering\footnotesize
\caption{Response-field consistency and sensing stability ($10^{-3}$ m/s$^2$; $\downarrow$ lower is better).}
\label{tab:evidencelocalrate}
\setlength{\tabcolsep}{3pt}
\begin{tabular}{lrrrr}\toprule
Estimation variant & 1 Hz & 5 Hz & 25 Hz & 125 Hz\\\midrule
\multicolumn{5}{l}{\emph{Clean acquisition agreement: $D_{\mathrm{rate}}$}}\\
Finite difference & 17.036 & 4.730 & 1.508 & 0$^{\dagger}$\\
Ordinary integral & \textbf{13.194} & \textbf{2.683} & \textbf{0.496} & 0$^{\dagger}$\\
\textbf{Weak (ours)} & 23.892 & 5.040 & 0.738 & 0$^{\dagger}$\\
\textbf{Exp. weak (ours)} & 26.627 & 5.158 & 0.773 & 0$^{\dagger}$\\
\midrule
\multicolumn{5}{l}{\emph{Independent error: $D_{\mathrm{ref}}$}}\\
Finite difference & 71.728 & 59.577 & 75.006 & 117.953\\
Ordinary integral & \textbf{66.087} & 55.109 & 49.880 & 48.955\\
\textbf{Weak (ours)} & 67.782 & \textbf{39.590} & \textbf{19.928} & \textbf{9.865}\\
\textbf{Exp. weak (ours)} & 69.861 & 40.661 & 20.500 & 10.084\\
\midrule
\multicolumn{5}{l}{\emph{Correlated error: $D_{\mathrm{ref}}$}}\\
Finite difference & 78.293 & 67.910 & 68.621 & 67.744\\
Ordinary integral & 71.497 & 65.583 & 65.842 & 65.732\\
\textbf{Weak (ours)} & \textbf{70.151} & \textbf{60.553} & \textbf{60.691} & \textbf{60.684}\\
\textbf{Exp. weak (ours)} & 72.057 & 61.525 & 61.720 & 61.787\\
\bottomrule\end{tabular}

{\footnotesize $^{\dagger}$Reference zeros are unranked.}
\end{table}

\begin{table}[!b]
\centering\footnotesize
\caption{ACC/manual separation using PI-Koopman residuals.}
\label{tab:claimbehavior}
\setlength{\tabcolsep}{3pt}
\begin{tabular}{lcccc}\toprule
& \multicolumn{2}{c}{Median $S_r$ ($10^{-3}\,\mathrm{m^2/s^4}$)} & \multicolumn{2}{c}{ $\uparrow$ Higher is better}\\
Model & ACC & Manual & AUC & BA\\\midrule
PI-Koopman-CTH & 2.248 & 25.366 & 0.938 & 0.854\\
PI-Koopman-IDM & 1.931 & 20.802 & 0.931 & 0.799\\
\bottomrule\end{tabular}
\end{table}

\begin{table*}[!t]
\centering\footnotesize
\caption{Target adaptation and 5-s forecast time (ms; $\downarrow$ lower is better).}
\label{tab:compute}
\setlength{\tabcolsep}{9pt}
\begin{tabular}{lccc}\toprule
Method & \shortstack{Apple M4 Pro} & \shortstack{AMD Threadripper PRO 7985WX} & \shortstack{NVIDIA RTX 6000 Ada with CUDA}\\\midrule
\multicolumn{4}{l}{\textit{PI identification and forecast (median)}}\\[2pt]
\textbf{PI-Koopman-CTH (ours)} & \textbf{11.00} & \textbf{16.64} & \textnormal{N/A}\\
\textbf{PI-Koopman-IDM (ours)} & 17.28 & 25.96 & \textnormal{N/A}\\
\midrule
\multicolumn{4}{l}{\textit{Neural 400-update workload and forecast}}\\[2pt]
DNN & 573 & 930 & 2230\\
LSTM & 909 & 1358 & 3380\\
\bottomrule\end{tabular}
\end{table*}

\begin{figure*}[!t]
\centering\includegraphics[width=\textwidth]{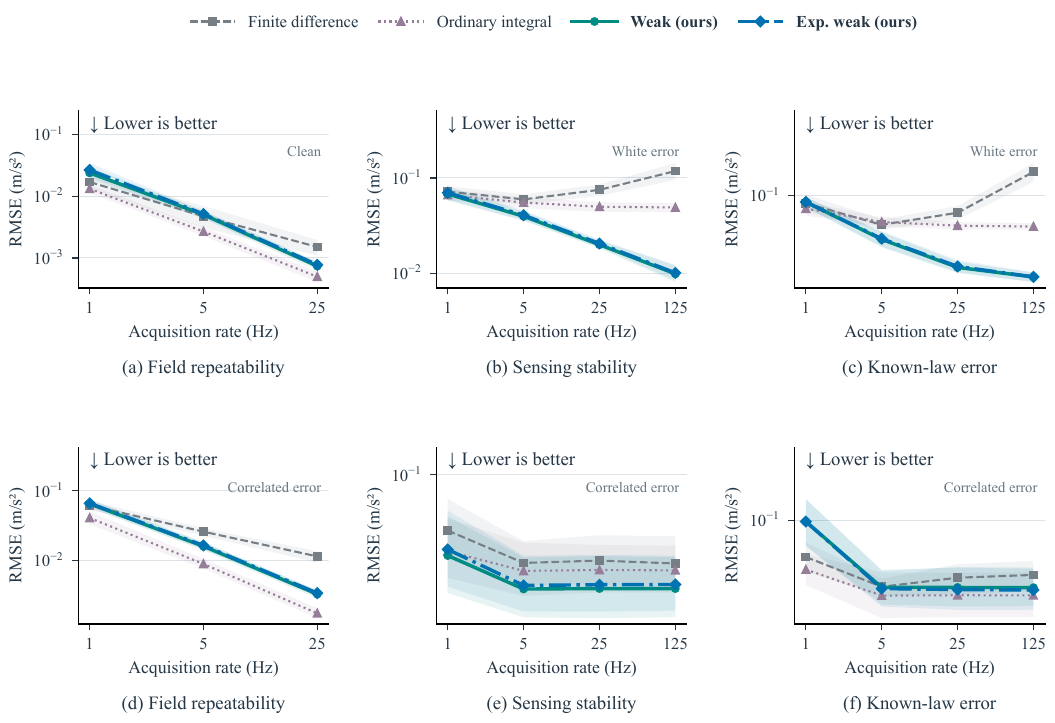}
\caption{Response-field consistency and accuracy under paired sensing interventions.}
\label{fig:ratefield}
\end{figure*}

\subsection{Response stability across sensing conditions}
\label{sec:rateresults}
The CATS-CF intervention shows how the same estimation mechanism uses real high-rate vehicle measurements. Under added independent errors at 125 Hz, compact weak estimation gives mean field deviation $D_{\mathrm{ref}}=0.00986$ m/s$^2$, against 0.11795 for finite differences and 0.04895 for ordinary integration (Table~\ref{tab:evidencelocalrate}). The corresponding perturbation reductions are 91.6\% and 79.8\%, each measured against the estimator's own clean high-rate fit. The paired 95\% intervals for weak-minus-FD and weak-minus-integral deviations are $[-0.1323,-0.0858]$ and $[-0.0446,-0.0337]$ m/s$^2$.

High-rate observations strengthen this averaging effect. Compact weak deviation under independent errors falls from 0.03959 m/s$^2$ at 5 Hz to 0.01993 at 25 Hz and 0.00986 at 125 Hz, with positive 95\% intervals for the reduction relative to finite differences and ordinary integration at all three rates. With correlated errors, compact weak estimation gives the lowest mean perturbation at every tested rate, with 5-125 Hz values near 0.061 m/s$^2$. The distinct profiles show how temporal correlation affects the information gained from denser sampling.

Clean compact weak fields at 25 and 5 Hz differ from their 125-Hz fits by 0.000738 and 0.00504 m/s$^2$, respectively. A matched exact-law control connects this real-data stability measure to reconstruction accuracy. Under the same independent-error process and 8-s calibration at 125 Hz, compact weak RMSE is 0.03788 m/s$^2$, compared with 0.13265 for FD and 0.06911 for integration, a 71.4\% reduction from FD. Figure~\ref{fig:ratefield} presents the paired rate, noise, and exact-law results with 95\% condition/object bootstrap bands. OpenACC establishes predictive performance across production platforms, while the CATS-CF measurement chain exposes how sampling and temporal estimation shape the recovered response.

\begin{figure*}[!t]
\centering\includegraphics[width=\textwidth]{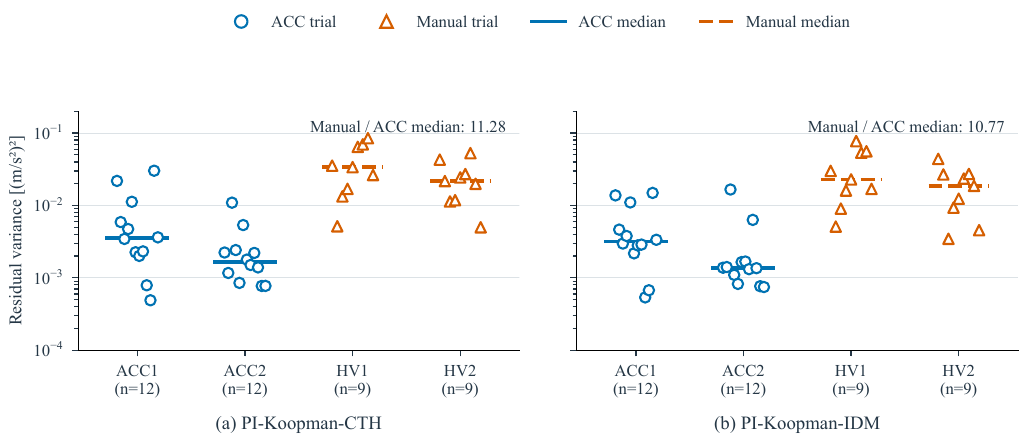}
\caption{PI-Koopman residual variability across following modes.}
\label{fig:pi-residual-modes}
\end{figure*}

\subsection{Response consistency for ACC/manual recognition}
\label{sec:behavior}
PI-Koopman exposes a marked difference in how consistently a fixed local response explains ACC and manual following. Figure~\ref{fig:pi-residual-modes} shows the residual variance of every encounter under both physical priors. For PI-Koopman-CTH, the median is 0.00225 $(\mathrm{m/s^2})^2$ for ACC and 0.02537 $(\mathrm{m/s^2})^2$ for manual driving, a factor of 11.28. PI-Koopman-IDM gives a corresponding factor of 10.77. The identified PI response supplies a behavioral signature through its residuals.

A single threshold on PI-Koopman-CTH residual variance achieves 85.4\% run-disjoint balanced accuracy, recognizing 21 of 24 ACC encounters and 15 of 18 manual encounters. Its AUC is 0.938 (Table~\ref{tab:claimbehavior}), and PI-Koopman-IDM gives 79.9\% balanced accuracy and 0.931 AUC. On the common-ACC-leader subset, PI-Koopman-CTH retains 83.3\% balanced accuracy and 0.931 AUC.

The PI-Koopman identification framework also supports a lightweight following-mode decision. Repeated CATS-CF settings and drivers make this connection observable across independent runs.

\subsection{Computation for frequent target adaptation}
\label{sec:computation}

PI-Koopman supports frequent response updates on both CPU systems (Table~\ref{tab:compute}). Fitting and a 5-s forecast take median 11.00/17.28 ms for the CTH/IDM branches on M4 Pro and 16.64/25.96 ms on the Threadripper PRO 7985WX CPU, within a 100-ms observation cycle. Timings include physical calibration and weak regression over six ACC records per dataset, all three histories, one warmup, and seven repetitions.

A fixed 400-update, 1-s-rollout DNN/LSTM workload measures neural adaptation cost across the three environments. It uses one OpenACC target and averages the three history-budget medians, each based on three repetitions. Totals include initialization, transfers, fitting, and forecasting. CPU execution uses one thread, and CUDA uses synchronized timing on a shared RTX 6000 Ada under concurrent load. These measured workloads show how compact response fitting supports frequent encounter-specific updates.

\section{Conclusion}
PI-Koopman combines a calibrated physical law with a compact generator correction learned through weak temporal equations. This representation identifies a queryable longitudinal response from short encounters. Across histories of 5, 7, and 11~s, the CTH branch reduces mean speed prediction error by 16.0\% on OpenACC and 15.4\% on CATS-CF relative to the best neural means. The PI-Koopman branches achieve the lowest mean speed and gap errors in both datasets. Known-law reconstruction and paired CATS-CF measurements demonstrate how smooth temporal testing improves response recovery from noisy observations. The same PI formulation supports following-mode recognition through residual consistency, with an approximately eleven-fold ACC/manual variance separation and 85.4\% run-disjoint balanced accuracy. Millisecond-scale CPU fitting and forecasting connect this data efficiency to frequent response updates for downstream traffic applications.

\IEEEtriggeratref{29}
\bibliographystyle{IEEEtran}
\bibliography{references}
\end{document}